\documentclass[a4paper]{article}
\usepackage{graphicx}
\usepackage[square,sort&compress]{natbib}
\usepackage{overpic}
\usepackage{amsmath}
\usepackage{hyperref}
\usepackage{a4wide}
\begin{document}

\title{A template method for measuring the iron spectrum in cosmic rays with Cherenkov telescopes}
\date{}
\author{Henrike Fleischhack\footnote{DESY, Platanenallee 6, 15738 Zeuthen. \url{henrike.fleischhack@desy.de} } ~for the VERITAS Collaboration\footnote{\url{http://veritas.sao.arizona.edu/}} }

\maketitle

\begin{abstract}
Understanding the sources, acceleration mechanisms, and propagation of cosmic rays is an active area of research in astro-particle physics. Measuring the spectrum and elemental composition of cosmic rays on earth can help solve this question. IACTs, while mainly used for $\gamma$-ray astronomy and indirect searches for dark matter, can make an important contribution here. In particular, they are able to distinguish heavy nuclei in cosmic rays from protons and lighter nuclei by exploiting the direct Cherenkov light emitted by charged particles high in the atmosphere. In this paper, a method to reconstruct relevant properties of primary cosmic ray particles from the Cherenkov light emitted by the primary particles and the air showers induced by them will be presented.
\end{abstract}

\section{Introduction}
Cosmic rays (CRs), charged particles of extraterrestrial origin impinging on the earth's atmosphere, were discovered more than one hundred years ago. Today, we know that they mostly consist of protons and other fully ionized nuclei, and (to a lesser extent) of electrons, positrons, and anti-protons. From a few GeV to several EeV, their energy spectrum can be described by a power law with two breaks: the ``knee'' in the PeV range and the ``ankle'' in the EeV range. The flux of lower energy particles is attenuated by the Sun's and the Earth's magnetic fields and hence undergoes seasonal changes.

\begin{figure}[htb]
\begin{center}
\includegraphics[width=8cm, trim=19cm 0.7cm 10cm 1.5cm, clip]{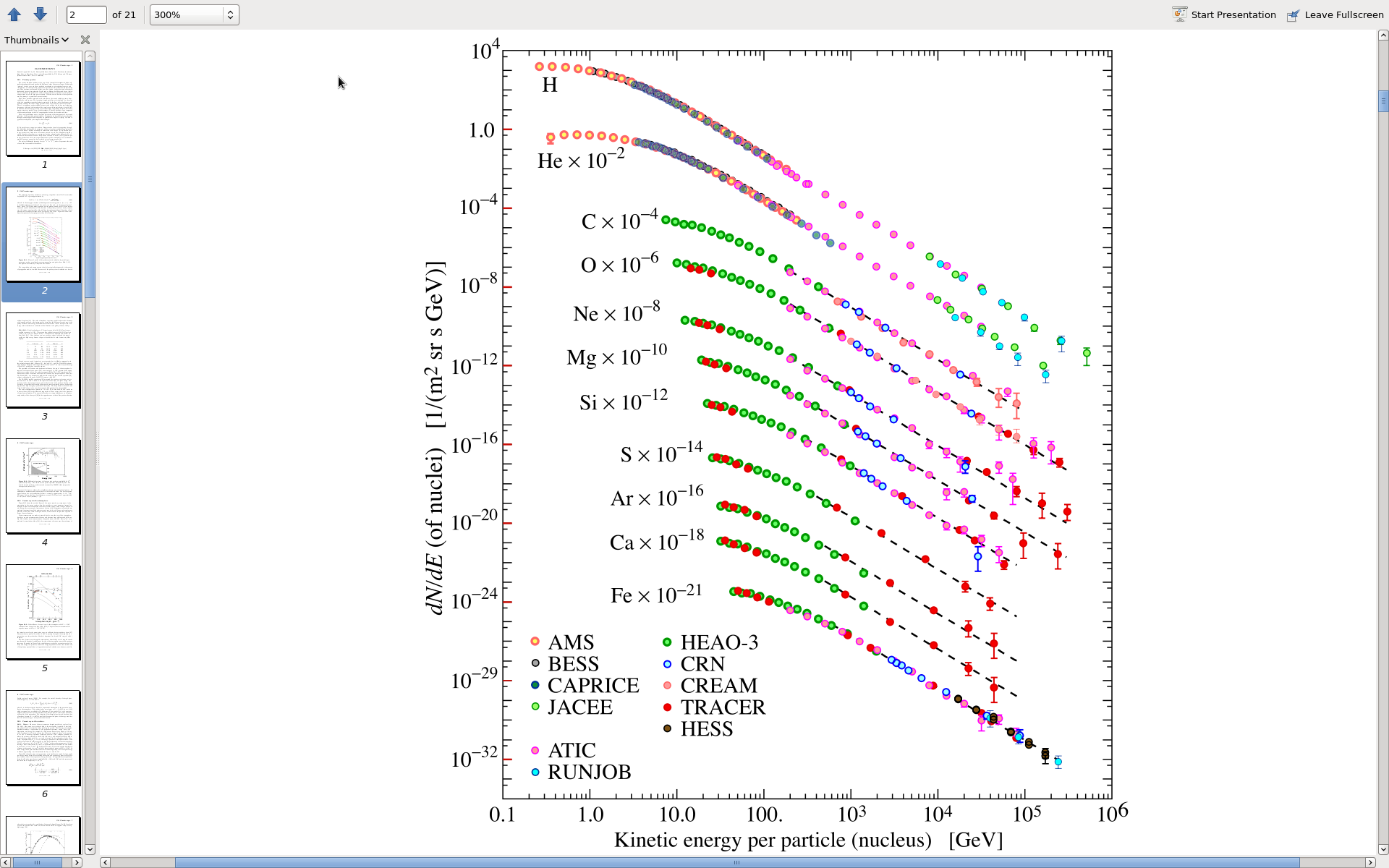}
\caption{\label{pdgspectrum}Energy spectra of various elements in cosmic rays, measured by different experiments over a large energy range. From \cite{PDG-2012}.}
\end{center}
\end{figure}

Many possible acceleration sites for cosmic rays have been identified, for example supernova remnants for galactic cosmic rays and gamma-ray bursts or active galactic nuclei for extra-galactic cosmic rays. However, it is still unknown by what mechanism the majority of cosmic rays are accelerated and what causes the change in spectral slope referred to as the ``knee''. Cosmic rays are charged particles, and are thus deflected by (inter-)galactic magnetic fields. Hence, their arrival direction does not point directly to the sources. We can only study the sources and acceleration mechanisms of cosmic rays by indirect means: Either by detecting neutral particles (neutrinos or $\gamma$-rays) which are produced at the sites of cosmic ray acceleration, or by precisely measuring the spectrum and composition of cosmic rays and comparing to theory predictions of cosmic ray acceleration and propagation. For an overview of such models and comparison of available data, see \cite{knee}. Even at energies below the knee, measurements of the flux and spectrum of different elements in cosmic rays can help constrain those models.

As the energies of cosmic rays impinging on earth span several orders of magnitude, a range of different techniques is employed to detect cosmic rays and investigate their properties. Some of the resulting energy spectra for the different elements that make up cosmic rays are shown in figure \ref{pdgspectrum}.

\subsection{Direct Detection Experiments}
Cosmic rays in the MeV to TeV range are detected directly at the top of the atmosphere by balloon- or satellite-borne detectors. As charged particles, they will deposit energy inside the detector by ionization and bremsstrahlung. Energy and charge of the primary particle can be inferred from the deposited energy and/or the shape of the particle's track, e.g. bending in a magnetic field, or deflection due to multiple scattering off atomic nuclei. The collection area is constrained by the size/weight limits of the balloon/satellite and is usually below $1\,$m$^2$. At energies above a few hundred GeV, the energy resolution of direct experiments tends to be limited by their size. Due to the power-law distribution of cosmic ray energies, they tend to  be statistics-limited at higher energies as well. For an example, see \cite{ahn2009}.

\subsection{Indirect Detection with EAS Arrays}
Cosmic rays with energies in the GeV range or higher will cause a cascade of particles (composed mostly of protons, neutrons, electrons, positrons, pions, muons and neutrinos) to be produced upon interaction with the atmosphere. For primary particles of hundreds of TeV and higher, parts of this so-called ``extended air shower'' of particles (muons, electrons) can be detected on the Earth's surface. By measuring the density and arrival times of these particles, energy and direction of the primary particles can be inferred. Charge measurements are difficult with this type of experiment. The composition of the primary particles can, for example, be inferred on a statistical basis from measurements of the height of the shower maximum, or from the particle content of the air shower reaching the ground. Both methods rely heavily on numerical simulations of air showers.

The footprint of an air shower can be several kilometers in diameter at the surface. A large detection area is needed to make up for the low flux of cosmic rays expected at such high energies. Experiments for the detection of air showers thus consist of an array of detector stations, hundreds of meters apart. Each station detects charged particles by the energy they deposit in the detector. These arrays can have detection areas up to $10^{12}\,$m$^2$. The surface detectors can be supported by arrays of radio antennae and/or fluorescence detectors, which measure radiation emitted by the charged component of the shower in the atmosphere. For examples, see \citet{abbasi2010, cazon2012}.

\subsection{Indirect Detection with IACTs}
Another way to detect extended air showers on the ground is by looking for the Cherenkov radiation emitted by the charged particles inside an air shower whenever they are faster than the (local) speed of light. Cherenkov radiation is highly beamed. The resulting light cone (footprint of the Cherenkov light on the ground) has a diameter of a few hundred meters; the size of the footprint does not depend much on the primary particle's energy. 

Imaging Air Cherenkov Telescopes are a class of instruments that exploit this measurement technique. They do not rely on the charged  component of the shower reaching the ground, but are able to self-trigger on the Cherenkov light. With their cameras, each made of several hundred or thousand sensitive light detectors (e.g. PMTs or silicon detectors), they can image entire air showers in Cherenkov light. From the size, shape, and position of the images in the camera, the energy and direction of the primary particles can be inferred. Arrays of several IACTs may be used together in a stereoscopic mode to improve the reconstruction.

IACTs are mostly used for $\gamma$-ray astronomy, detecting air showers produced by TeV-range photons. In that case, cosmic rays make up the main background. However, they can also be used to study properties (energy spectrum and composition) of cosmic rays.

With their effective detection area of about $10^4\,$m$^2$, IACTs are able to cover intermediate energies (TeV range) to fill the gap between direct detection experiments and air shower arrays, cf. \cite{rolfpaper}. In contrast to air shower arrays, it is possible to measure the charge of the primary particle on an event-to-event basis. 

\subsection{Direct Cherenov Technique}
Charged primary particles can radiate direct Cherenkov (DC) light even before starting a shower, if their velocity is high enough. This DC light can be detected with IACTs. It is very concentrated in the camera (inside one pixel in current IACTs), and located at the front of the shower image. The intensity of the DC light is proportional to the square of the charge of the primary particle, making it useful for the separation of heavy and light nuclei. At the same time, the Cherenkov light from the shower may be used to reconstruct energy and direction of the primary particle. The existence of DC light can in particular be used to select showers induced by heavy primary particles (in particular iron) in an energy range of about $10\,$TeV to several $100\,$TeV. The higher the energy, the less dense the atmosphere needs to be for the primary to emit Cherenkov light. At low energies, the particle needs to penetrate very deeply into the atmosphere to reach layers that are dense enough for Cherenkov light to be emitted. Below $10\,$TeV or so, the probability for the particle traveling that far in the atmosphere without interaction is very low. The higher the primary particle's energy, the higher in the atmosphere it starts emitting DC light. However, at higher energies above hundreds of TeV, the light emitted by the shower starts to drown out the DC contribution, making it impossible to infer the charge of the primary. cf. \cite{rolfpaper, Kieda}.

Most showers detected by IACTs are induced by cosmic rays. Thus, any study of cosmic rays using IACT data can exploit a large existing data set (typically up to 1000 hours per year of operation).Showers induced by heavy nuclei (in particular iron) can be separated from light nuclei and protons due to the extra component of the DC light. In the following, iron will be used as a representative for heavy nuclei as it is the most abundant element in cosmic rays with $Z>20$. Lighter elements such as helium, carbon and oxygen do appear more abundantly among cosmic rays. However, the intensity of the direct Cherenkov light is about ten times lower for oxygen compared to iron, making it difficult to distinguish showers induced by light elements from proton-induced showers using this technique.

Measuring the iron flux and spectrum this way is beneficial because the systematic uncertainty is complementary to other cosmic ray experiments (for IACTs, the atmosphere is the largest source of systematic uncertainty).

\subsection{The VERITAS Experiment}
VERITAS\footnote{\url{http://veritas.sao.arizona.edu/}} (Very Energetic Radiation Imaging Telescope Array System) \cite{veritas} is an array of four IACTs, located in southern Arizona, USA. It has been operating since 2007. Each telescope has a set of tessellated mirrors with a total mirror diameter of about $12\,$m$^2$. Each camera consists of about 500 ``pixels'' (PMTs), with a total field of view of about 3.5 degrees. It is sensitive to gamma-ray induced showers from 80 GeV to tens of TeV.

Over its lifetime, the telescope has seen several upgrades: one of the telescopes was moved in 2009 to increase the instrument's effective area; the PMTs were upgraded in 2012 to increase sensitivity for fainter showers. Science topics studied with VERITAS include gamma-ray astronomy, dark matter searches, and astro-particle physics.

\section{Data analysis with IACTs}
\subsection{Standard geometrical analysis method}
Cherenkov telescopes record signals from air showers. From those, properties of the primary particle have to be reconstructed. For gamma-ray astronomy, the relevant properties are energy, arrival direction, and particle class (photon or cosmic ray).

The standard geometrical reconstruction proceeds as follows. First, the signal in each pixel is integrated. The pedestal is subtracted and the signal is corrected for the gain. A cleaning is performed to reduce contributions from the night sky background.  

Then, a moments analysis is performed on the surviving image pixels and the so-called `Hillas parameters' \cite{hillas} are calculated. To first order, gamma-ray induced showers develop symmetrically. Thus, the major axis of the image points back to the source of the gamma-ray. By combining information from two or more telescopes, the arrival direction of the gamma ray and the position of the shower core on the ground can be determined. Once the distance between shower core and telescope are known, lookup tables can be used to estimate the primary particle's energy from the total amount of light in the camera. Then, the second moments (width and length) are compared to the expected values for a gamma-induced shower of that energy. Showers induced by cosmic rays tend to be wider and less even than gamma-ray induced showers. Thus, the width and length of the images are a good discriminator between gamma-ray and cosmic-ray induced showers \cite{fegan}.

\subsection{The Template Likelihood analysis method}
\begin{figure}[tb]
\begin{center}
\includegraphics[width=20pc]{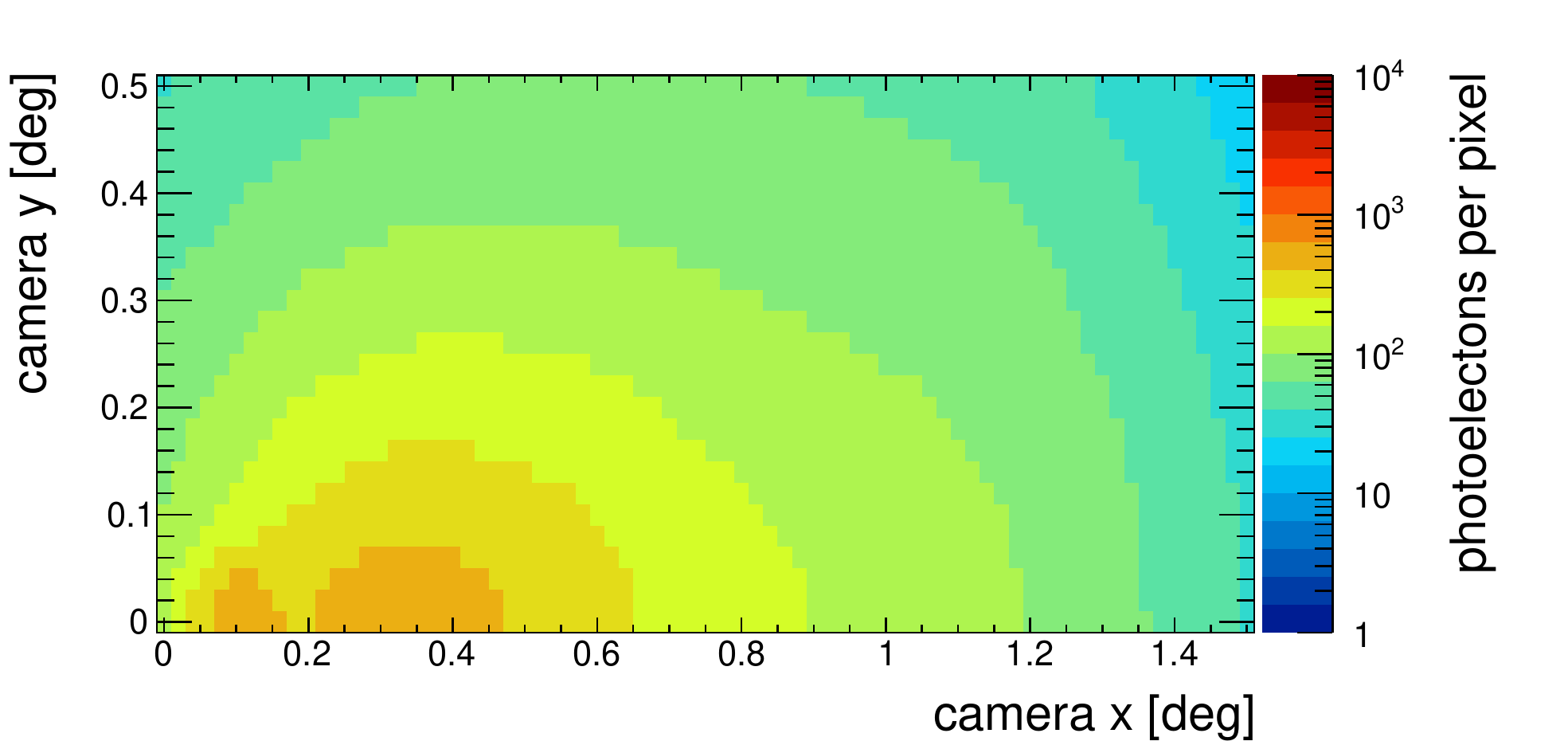}
\caption{\label{template}Average light distribution in the camera for $E=30\,$TeV, $50\,$m distance between the detector and shower core, $h=33\,$km. The coordinate system is centered on the primary particle's arrival direction. The x-axis is defined by the long image axis. Note the contribution from direct Cherenkov light at around 0.1 degrees from the primary's direction, clearly distinguishable from the broader contribution from the Cherenkov light emitted by the shower particles.}

\end{center} 
\end{figure}
The geometrical analysis described above works very well to reconstruct gamma-induced showers, but it does not take all available information into account. Other, more advanced techniques have been developed in the past years. One of them is the template likelihood method, which naturally takes the information from each pixel into account. The idea is as follows: Given a model which describes the photon distribution in the camera, depending on some properties of the primary particle, we can model the probability distribution of collected charge per pixel given those parameters as well as the detector response and its uncertainty. This gives us a likelihood function, which can be maximized to estimate the parameters for a given shower by fitting camera images to model predictions. The goodness of fit can be used to discriminate between `signal' and `background' events. This type of likelihood fitting has been used by similar experiments for the reconstruction of gamma-ray induced showers, see for example \cite{cattemplate}. It is particularly useful to discriminate between iron-induced showers, which can have a contribution from direct Cherenkov light, and showers induced by protons and light elements, where the direct Cherenkov light is not detectable.

In this case, the model parameters are the primary particle's energy $E$, direction in the camera ($X_s$, $Y_s$), height of first interaction $h$ and position of shower core on the ground ($X_p$, $Y_p$). The light distribution on the camera is modeled using Monte Carlo simulations of iron-induced showers at a fixed `grid' in these parameters, interpolating between them to predict the light distribution for arbitrary parameter values. CORSIKA \cite{corsika} is used to simulate air showers including Cherenkov light emission, and the grisudet package \footnote{\url{http://www.physics.utah.edu/gammaray/GrISU/}} for ray-tracing in the telescope.

An example for the predicted average light distribution in the camera can be seen in figure \ref{template}. In this example, the contributions from the shower light and from the direct Cherenkov light form two bumps that can be clearly separated by eye. The direct Cherenkov light is emitted at a very small angle high up in the atmosphere, and confined to a region in the camera that is less than a pixel in size. Images of showers induced by protons, light elements, and heavy elements like iron look very similar, but the direct Cherenkov contribution, which is proportional to the square of the particle's charge, can be used to distinguish between them. Hence, it is necessary to take information from all the pixels into account. The likelihood method described above does just that; an image from a proton-induced shower will not have a direct Cherenkov contribution and will thus have a worse goodness-of-fit value than an iron-induced shower when using templates generated for iron showers.
%`Signal' events are iron-induced showers, the `background' is made up of the rest of the cosmic rays, mostly protons and helium. Gamma-rays are very rare, even for very strong sources, less than one in a thousand triggered events is caused by a gamma-ray induced shower.
%\clearpage

The probability distribution per pixel is given by

\begin{align*}
P(q|s,\sigma_p, \sigma_e, \sigma_s) =& \int \mathrm{d}\mu ~ Gauss(\mu|s, \sigma_s)\cdot \sum\limits_n Poisson(n|\mu) \cdot Gauss(q|n,\sqrt{\sigma_p^2 + n\sigma_e^2}) \\
\approx& ~ Gauss(q|s,\sigma) ~~\mathrm{with} ~~ \sigma = \sqrt{\sigma_p^2+s\sigma_e^2+\sigma_s^2}
\end{align*} 
%\clearpage
Where
\begin{itemize}
\item $q$ is the pedestal-corrected integrated charge (converted to units of photo-electrons [p.e.]).
\item $s=s(E,h,X_p,Y_p,X_s,Y_s)$ and $\sigma_s=\sigma_s(E,h,X_p,Y_p,X_s,Y_s)$ are the predicted average number of photo-electrons in each pixel and its uncertainty. 
\item $\mu$ is the predicted number of photo-electrons in a given shower, assumed to follow a Gaussian distribution with mean $s$ and width $\sigma_s$.
\item $n$ is the number of photo-electrons, assumed to follow a Poisson distribution with mean $\mu$. 
\item $\sigma_e^2$ is the variance of the signal (integrated charge) from a single photo-electron.
\item $\sigma_p^2$ is the variance of the pedestal.
\end{itemize}
\clearpage
The log-likelihood function to be minimized is then given by
\begin{align*}
-\ln L = - \sum\limits_{\mathrm{pixel}~i} \ln L_i = \sum\limits_{\mathrm{pixel}~i} \ln P(q_i,s_i,\sigma_i). 
\end{align*}

We define the goodness-of-fit function as follows:
\begin{align*}
G=\frac{\sum_i \ln L_i - \left\langle \ln P(q|s_i,\sigma_i)\right\rangle_q }{\sqrt{2\cdot NDF}}.
\end{align*}
It has been normalized so that it follows a distribution with a mean of zero and width of one for signal events and can thus be used to discriminate between signal and background.

\section{Method Validation}

\begin{figure}[t]
\begin{center}
\begin{minipage}[t]{15pc}
\begin{overpic}[width=15pc]{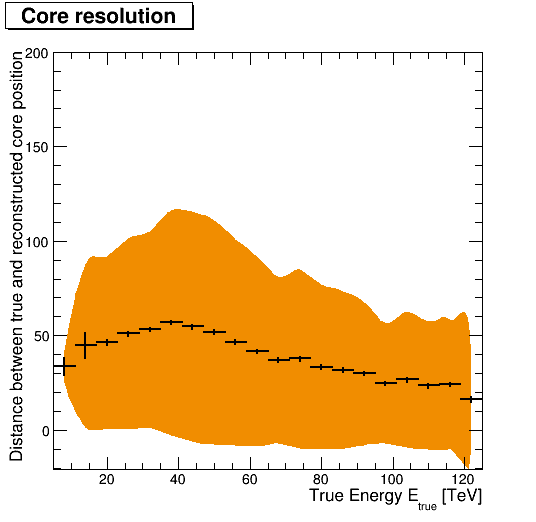}
\put( 50, 75) {\fbox{\textsf{\textbf{Preliminary}}}}
\put( 1.7, 87.8) {\tiny{\textsf{\rotatebox{90}{[m]}}}}
\end{overpic}
\caption{\label{corres}Distance between true and reconstructed shower core in $m$ vs true energy. The histogram gives the mean value in each energy bin, the shaded region corresponds to the 1-$\sigma$ width of the distribution.}
\end{minipage}\hspace{2pc}%
\begin{minipage}[t]{15pc}
\begin{overpic}[width=15pc]{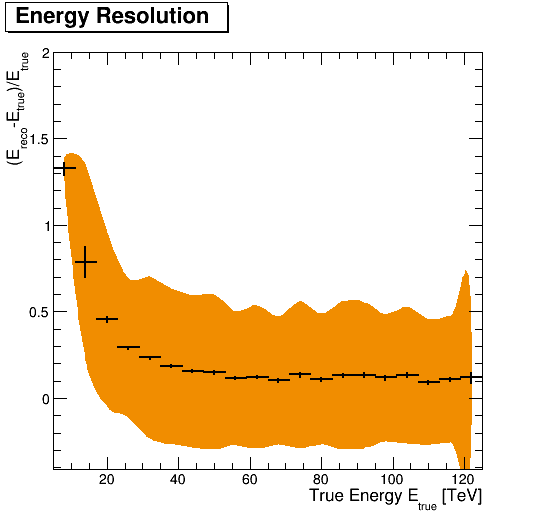}
\put( 50, 75) {\fbox{\textsf{\textbf{Preliminary}}}}
\end{overpic}
\caption{\label{Eres}Relative error in reconstructed energy vs true energy. The histogram gives the mean value in each energy bin, the shaded region corresponds to the 1-$\sigma$ width of the distribution.}
\end{minipage}
\end{center} 
\end{figure}

To test the method, a set of templates for showers from zenith and an energy range from $10$ -- $100\,$TeV was produced. A second set of showers was produced with random core positions, with the energies following a power law spectrum. This second set of showers was passed through a VERITAS detector simulation chain, again using the grisudet package. The resulting `fake data' files were analyzed using the likelihood fitting method described above. 

As a first check, the fit was performed using the true values of the parameters as starting values. In that case, the reconstruction worked very well. %, with a resolution of the shower core position of less than $10\,$m. 
However, using the results of the standard geometrical reconstruction as the starting values for the fit worsens the resolution considerably. For example, figures \ref{corres} and \ref{Eres} show the resolution of the reconstructed core position and reconstructed energy, both as a function of the (known) true energy. The main problem is that the geometrical reconstruction of the core position can give very bad results, making the fit unable to reconstruct the event properly.

\section{Conclusions and Outlook}
Measuring the flux and spectrum of heavy nuclei in TeV cosmic rays can provide useful insights into the acceleration and propagation of cosmic rays. However, this energy range is not well covered by direct detection experiments nor by air shower arrays. IACTs are sensitive especially to heavy nuclei like iron in that energy range. They are sensitive to direct Cherenkov light emitted by cosmic rays high in the atmosphere. A technique for reconstructing the energy of iron-induced showers imaged by air Cherenkov telescopes, and to separate these showers from showers induced by protons and lighter elements, has been introduced. This technique uses a likelihood fitting method to compare images of cosmic ray showers to predictions based on Monte Carlo simulations. Further work is needed to either improve the starting values used for the fit, or to make the fit itself more robust against bad starting values.

\section*{Acknowledgements}
This research is supported by grants from the U.S. Department of Energy Office of Science, the U.S. National Science Foundation and the Smithsonian Institution, by NSERC in Canada, by Science Foundation Ireland (SFI 10/RFP/AST2748) and by STFC in the U.K. We acknowledge the excellent work of the technical support staff at the Fred Lawrence Whipple Observatory and at the collaborating institutions in the construction and operation of the instrument. The VERITAS Collaboration is grateful to Trevor Weekes for his seminal contributions and leadership in the field of VHE gamma-ray astrophysics, which made this study possible.  

\bibliography{bib}

\end{document}